
\font\biggerbf=cmbx12 at 20.736truept
\font\bigrm=cmr12 at 17.28truept
\font\bigbf=cmbx12 at 17.28truept
 at 17.28truept
 at 17.28truept
\font\smallsl=cmsl8 at 8truept
\font\eightrm=cmr9
\font\eighti=cmmi9
\font\eightsy=cmsy9
\font\eightit=cmti9
\font\eightbf=cmbx9


\catcode`@=11 
\def\ten{\def\rm{\fam\z@\tenrm}
  \textfont\z@=\tenrm \scriptfont\z@=\sevenrm \scriptscriptfont\z@=\fiverm
  \textfont\@ne=\teni \scriptfont\@ne=\seveni \scriptscriptfont\@ne=\fivei 
  \textfont\tw@=\tensy \scriptfont\tw@=\sevensy \scriptscriptfont\tw@=\fivesy 
  \textfont\itfam=\tenit  \def\it{\fam\itfam\tenit }
  \textfont\bffam=\tenbf \scriptfont\bffam=\sevenbf
  \scriptscriptfont\bffam=\fivebf \def\bf{\fam\bffam\tenbf }
  \normalbaselineskip=12pt plus .1pt
  \normalbaselineskip\rm}
\def\eight{\def\rm{\fam\z@\eightrm}
  \textfont\z@=\eightrm \scriptfont\z@=\fiverm \scriptscriptfont\z@=\fiverm
  \textfont\@ne=\eighti \scriptfont\@ne=\fivei \scriptscriptfont\@ne=\fivei 
  \textfont\tw@=\eightsy \scriptfont\tw@=\fivesy \scriptscriptfont\tw@=\fivesy 
  \textfont\itfam=\eightit  \def\it{\fam\itfam\eightit }
  \textfont\bffam=\eightbf \scriptfont\bffam=\fivebf
  \scriptscriptfont\bffam=\fivebf   \def\bf{\fam\bffam\eightbf }
  \normalbaselineskip=11pt plus .1pt
  \normalbaselineskip\rm}
\catcode`@=12

\def\Color#1#2{#2}
\def\Red#1{\Color{0. 1.0 1.0 0.} {#1}}
\def\Blue#1{\Color{1.0 1.0 0.0 0.} {#1}}

\tolerance=5000
\clubpenalty=10000
\widowpenalty=10000
\emergencystretch=2pt
\hfuzz=6pt

\def\nn{\par\noindent\hangindent=.20truein \hangafter=1}
\def\pp{\par\noindent\hangindent=.125truein \hangafter=1}


\parskip = 6pt plus 1pt minus 1pt

\def\phrase #1 #2 #3
 {\line{\hfil\Blue{\bf #2}\hfil}
  \vskip -\baselineskip
  \line{\Red{\smallsl #1}\hfil\Red{\smallsl #3}}
  \nobreak
  \smallskip
  \nobreak
  \line{\hfil$\downarrow$\hfil}
  \smallskip
  \goodbreak
 \relax}

\def\phrasehead #1 #2
 {\line{\Red{\smallsl #1}\hfil\Red{\smallsl #2}}
  \smallskip
  \smallskip
 \relax}

\def\phrasefoot #1
 {\line{\hfil\Blue{\bf #1}\hfil}
  \smallskip
  \smallskip
 \relax}

\ten

\centerline{\biggerbf The evolution of galaxy formation}
\bigskip
\centerline{\bigrm Douglas Scott}
\smallskip
\centerline{University of British Columbia}
\bigskip
\bigskip
\bigskip

\centerline{\bigbf Abstract}
\medskip
\noindent
{\narrower\narrower\noindent
Our history of understanding galaxy formation could be traced through
the development of individual ideas.  A cynic might be tempted to suggest
that new catchphrases are developed at a faster rate than genuine
progress is made.
\par}
\bigskip
\bigskip
\bigskip

\centerline{\bigbf The story so far}
\medskip
\noindent
Galaxy formation is a complex subject, and the fundamental question ``how do
galaxies form?'' is poorly defined.  Galaxies do not ``form''
instantaneously, and there is no sharp dividing line between their creation
and the evolution of their properties.
Understanding how lumps of a specific size developed from
simplicity to complexity is clearly a challenging endeavour.
Different researchers approach this
from a variety of directions, and the history of progress on this topic
is replete with ideas that have temporarily captured the imagination of
astrophysicists.
This history can be followed through a series of the most important review
(or review-like) articles, including
Gamow (1948),
Hoyle (1953),
Sciama (1955),
Eggen, Lyden-Bell \& Sandage (1962),
Layzer (1964),
Peebles (1965),
Doroshkevich, Zeldovich \& Novikov (1967),
Larson (1969),
Harrison (1970),
Peebles (1974),
Jones (1976),
Gott (1977),
Rees (1978),
Efstathiou \& Silk (1983),
Blumenthal et al.\ (1984),
White (1995),
Baugh (2006) and
Benson (2010),
as well as
books by Longair (1998, 2008), Spinrad (2005) and Mo, Van den Bosch \&
White (2010).

However, instead of reviewing these reviews, I
present a brief timeline for the development of the concepts underlying
our current understanding of galaxy formation [1].  This is best told via
pieces of what was once called
``rhetorick'' and in the modern age might be known as
``catchphrases'', ``sound-bites'' or ``spin'' [2].  For each of these physical
ideas, I have tried to identify the originator as well as when the idea
caught on, as traced through the first paper found with the relevant
phrase in the title or abstract [3].  These entries are arranged in
chronological order of first appearance of each phrase, which in some
cases may be a surprisingly long time after the concept was first described.
No attempt has been made to measure how prevalent each idea has been, or
to track its rise and fall [4].  Imperfect as it may be, this
list provides an interesting history of ideas in this perplexing branch of
astrophysics [5].

\vfil\eject

\phrasehead{ORIGIN} {USAGE}
\smallskip
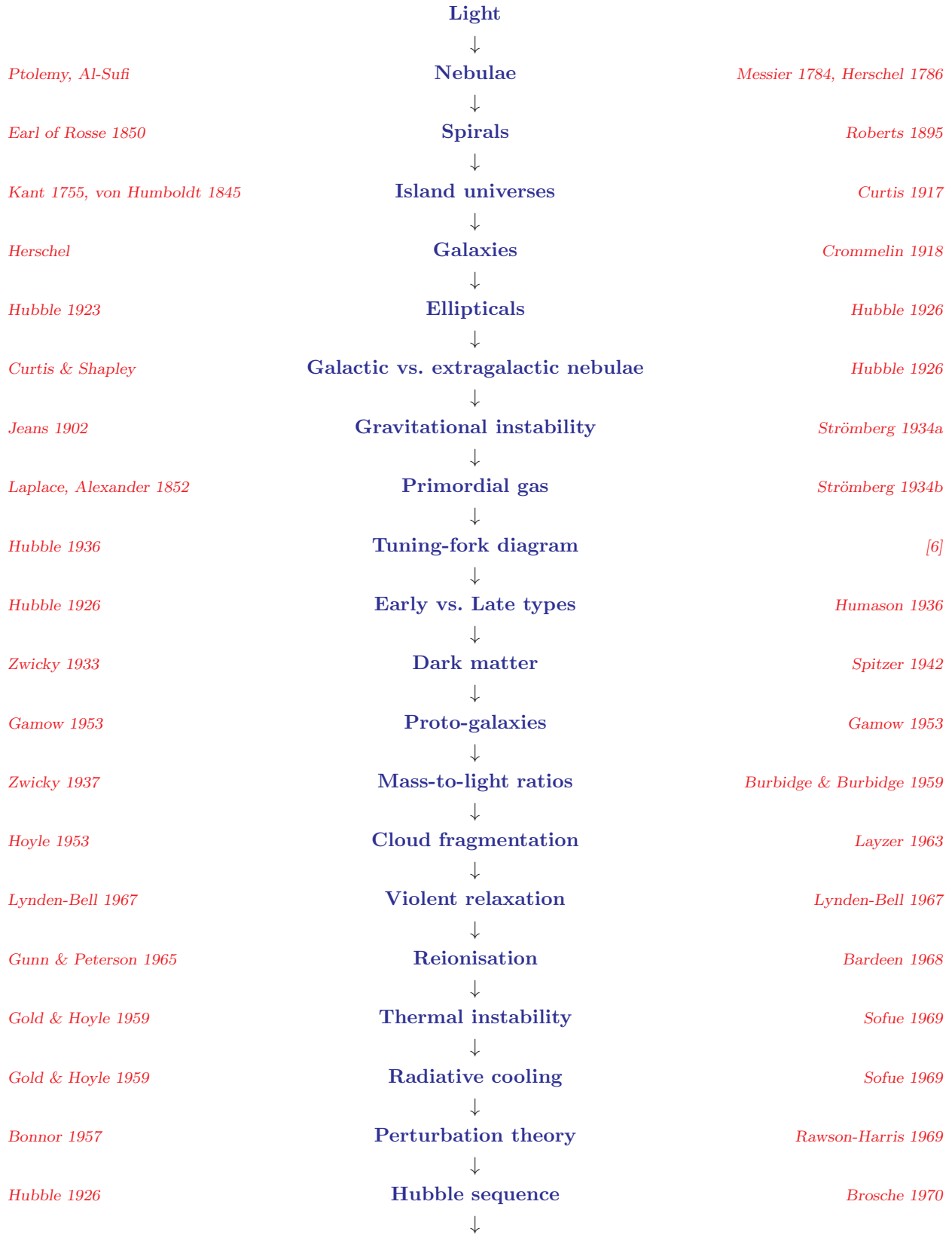
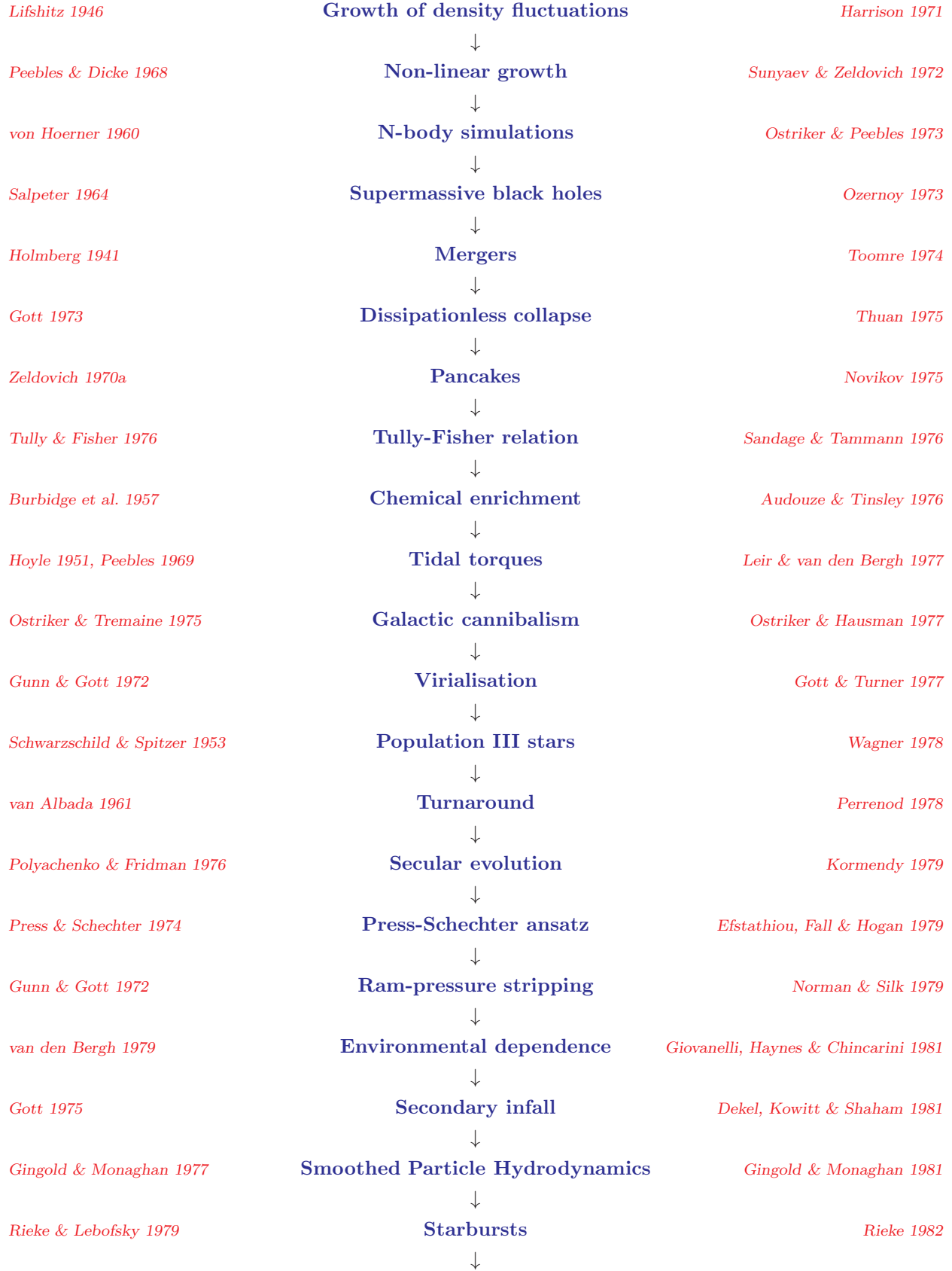
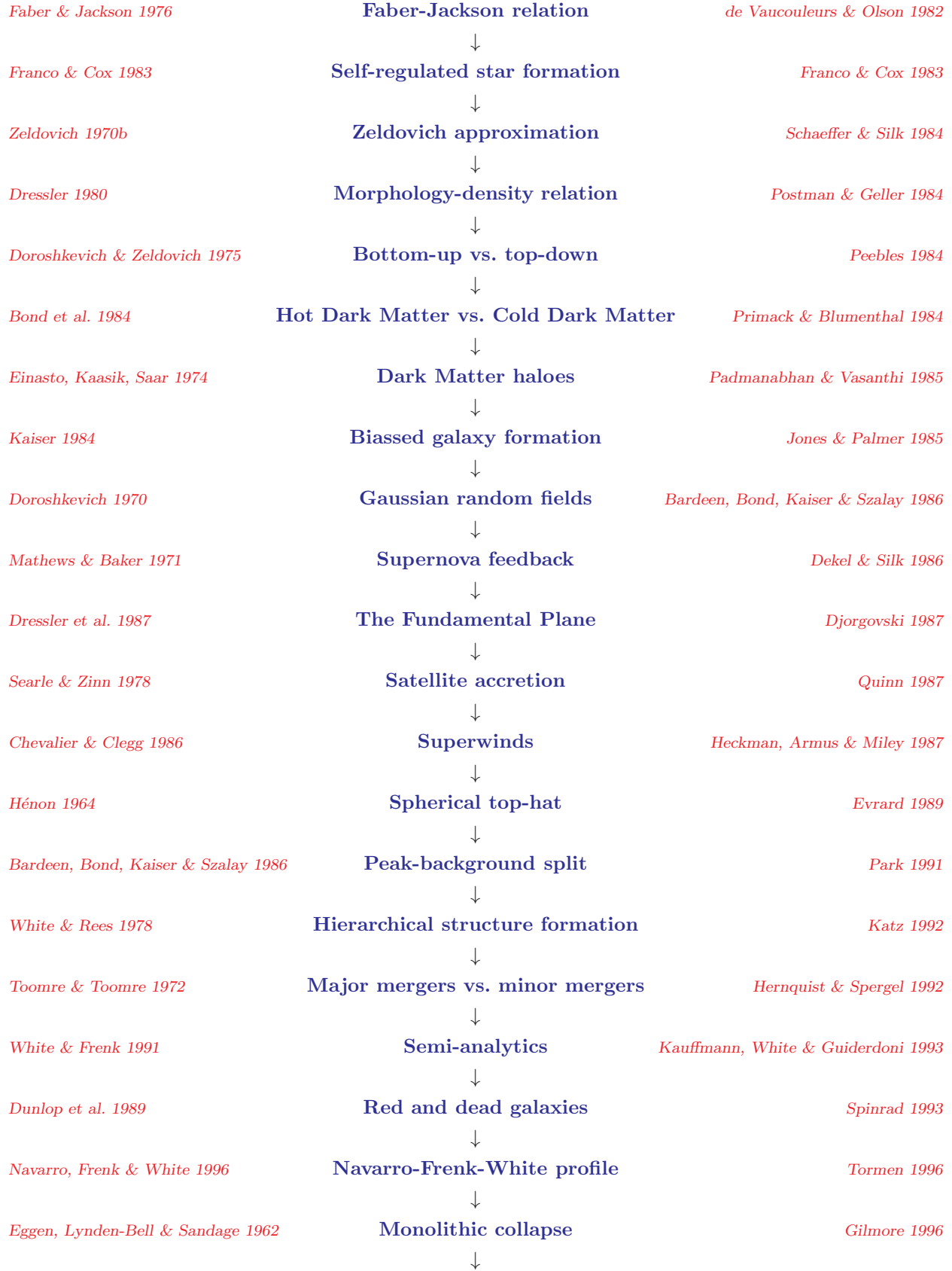
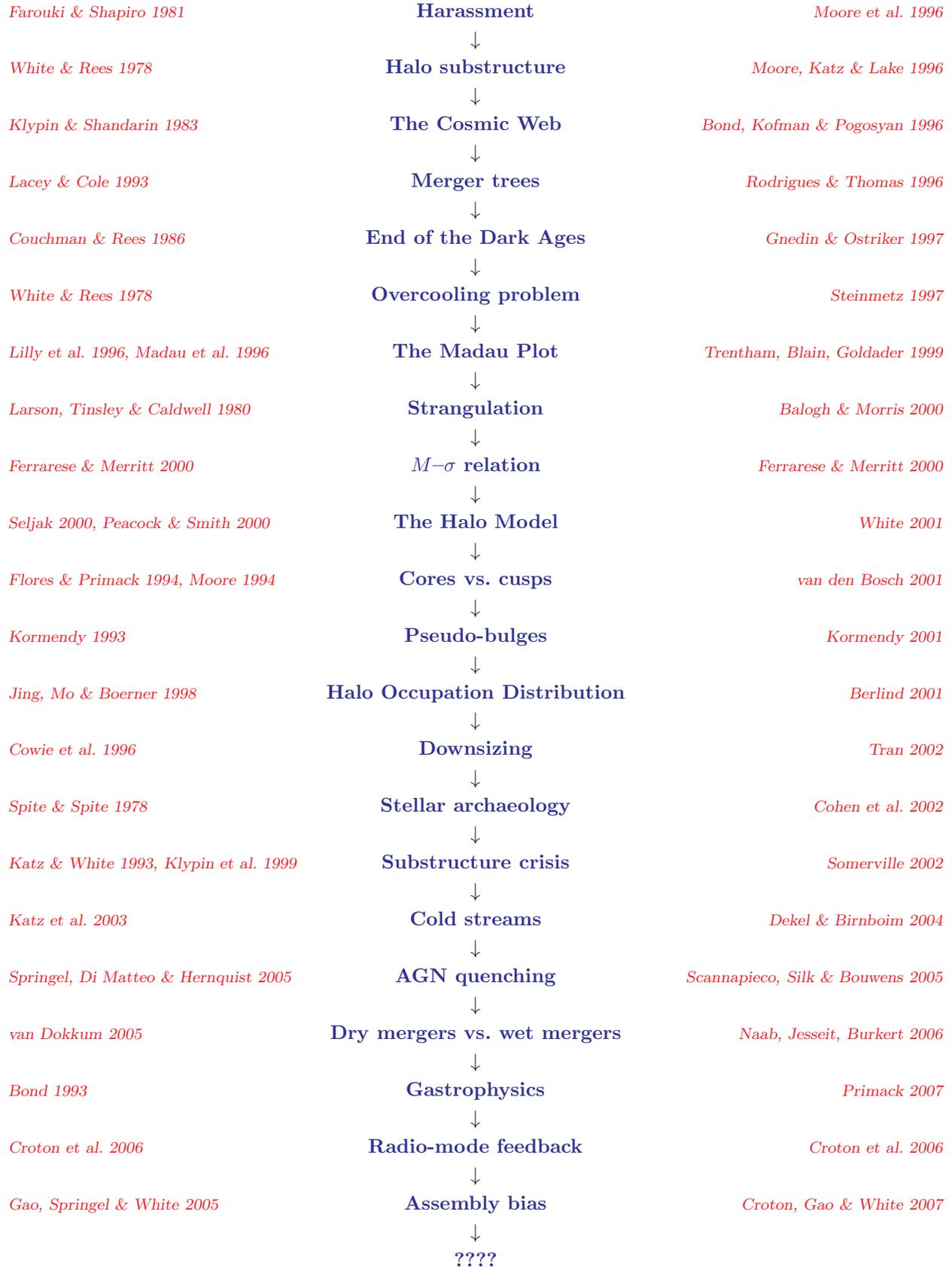
\phrase{} {Light} {}
\phrase{Ptolemy, Al-Sufi} {Nebulae} {Messier 1784, Herschel 1786}
\phrase{Earl of Rosse 1850} {Spirals} {Roberts 1895}
\phrase{Kant 1755, von Humboldt 1845} {Island universes} {Curtis 1917}
\phrase{Herschel} {Galaxies} {Crommelin 1918}
\phrase{Hubble 1923} {Ellipticals} {Hubble 1926}
\phrase{Curtis \& Shapley} {Galactic vs.\ extragalactic nebulae} {Hubble 1926}
\phrase{Jeans 1902} {Gravitational instability} {Str{\"o}mberg 1934a}
\phrase{Laplace, Alexander 1852} {Primordial gas} {Str{\"o}mberg 1934b}
\phrase{Hubble 1936} {Tuning-fork diagram} {[6]}
\phrase{Hubble 1926} {Early vs.\ Late types} {Humason 1936}
\phrase{Zwicky 1933} {Dark matter} {Spitzer 1942}
\phrase{Gamow 1953} {Proto-galaxies} {Gamow 1953}
\phrase{Zwicky 1937} {Mass-to-light ratios} {Burbidge \& Burbidge 1959}
\phrase{Hoyle 1953} {Cloud fragmentation} {Layzer 1963}
\phrase{Lynden-Bell 1967} {Violent relaxation} {Lynden-Bell 1967}
\phrase{Gunn \& Peterson 1965} {Reionisation} {Bardeen 1968}
\phrase{Gold \& Hoyle 1959} {Thermal instability} {Sofue 1969}
\phrase{Gold \& Hoyle 1959} {Radiative cooling} {Sofue 1969}
\phrase{Bonnor 1957} {Perturbation theory} {Rawson-Harris 1969}
\phrase{Hubble 1926} {Hubble sequence} {Brosche 1970}
\phrase{Lifshitz 1946} {Growth of density fluctuations} {Harrison 1971}
\phrase{Peebles \& Dicke 1968} {Non-linear growth} {Sunyaev \& Zeldovich 1972}
\phrase{von Hoerner 1960} {N-body simulations} {Ostriker \& Peebles 1973}
\phrase{Salpeter 1964} {Supermassive black holes} {Ozernoy 1973}
\phrase{Holmberg 1941} {Mergers} {Toomre 1974}
\phrase{Gott 1973} {Dissipationless collapse} {Thuan 1975}
\phrase{Zeldovich 1970a} {Pancakes} {Novikov 1975}
\phrase{Tully \& Fisher 1976} {Tully-Fisher relation} {Sandage \& Tammann 1976}
\phrase{Burbidge et al.\ 1957} {Chemical enrichment} {Audouze \& Tinsley 1976}
\phrase{Hoyle 1951, Peebles 1969} {Tidal torques} {Leir \& van den Bergh 1977}
\phrase{Ostriker \& Tremaine 1975} {Galactic cannibalism}
 {Ostriker \& Hausman 1977}
\phrase{Gunn \& Gott 1972} {Virialisation} {Gott \& Turner 1977}
\phrase{Schwarzschild \& Spitzer 1953} {Population III stars} {Wagner 1978}
\phrase{van Albada 1961} {Turnaround} {Perrenod 1978}
\phrase{Polyachenko \& Fridman 1976} {Secular evolution} {Kormendy 1979}
\phrase{Press \& Schechter 1974} {Press-Schechter ansatz}
 {Efstathiou, Fall \& Hogan 1979}
\phrase{Gunn \& Gott 1972} {Ram-pressure stripping} {Norman \& Silk 1979}
\phrase{van den Bergh 1979} {Environmental dependence}
 {Giovanelli, Haynes \& Chincarini 1981}
\phrase{Gott 1975} {Secondary infall} {Dekel, Kowitt \& Shaham 1981}
\phrase{Gingold \& Monaghan 1977} {Smoothed Particle Hydrodynamics}
 {Gingold \& Monaghan 1981}
\phrase{Rieke \& Lebofsky 1979} {Starbursts} {Rieke 1982}
\phrase{Faber \& Jackson 1976} {Faber-Jackson relation} {de Vaucouleurs \&
 Olson 1982}
\phrase{Franco \& Cox 1983} {Self-regulated star formation} {Franco \& Cox 1983}
\phrase{Zeldovich 1970b} {Zeldovich approximation} {Schaeffer \& Silk 1984}
\phrase{Dressler 1980} {Morphology-density relation} {Postman \& Geller 1984}
\phrase{Doroshkevich \& Zeldovich 1975} {Bottom-up vs.\ top-down} {Peebles 1984}
\phrase{Bond et al.\ 1984} {Hot Dark Matter vs.\ Cold Dark Matter}
 {Primack \& Blumenthal 1984}
\phrase{Einasto, Kaasik, Saar 1974} {Dark Matter haloes}
 {Padmanabhan \& Vasanthi 1985}
\phrase{Kaiser 1984} {Biassed galaxy formation} {Jones \& Palmer 1985}
\phrase{Doroshkevich 1970} {Gaussian random fields}
 {Bardeen, Bond, Kaiser \& Szalay 1986}
\phrase{Mathews \& Baker 1971} {Supernova feedback} {Dekel \& Silk 1986}
\phrase{Dressler et al.\ 1987} {The Fundamental Plane} {Djorgovski 1987}
\phrase{Searle \& Zinn 1978} {Satellite accretion} {Quinn 1987}
\phrase{Chevalier \& Clegg 1986} {Superwinds} {Heckman, Armus \& Miley 1987}
\phrase{H{\'e}non 1964} {Spherical top-hat} {Evrard 1989}
\phrase{Bardeen, Bond, Kaiser \& Szalay 1986} {Peak-background split}
 {Park 1991}
\phrase{White \& Rees 1978} {Hierarchical structure formation} {Katz 1992}
\phrase{Toomre \& Toomre 1972} {Major mergers vs.\ minor mergers}
 {Hernquist \& Spergel 1992}
\phrase{White \& Frenk 1991} {Semi-analytics}
 {Kauffmann, White \& Guiderdoni 1993}
\phrase{Dunlop et al.\ 1989} {Red and dead galaxies} {Spinrad 1993}
\phrase{Navarro, Frenk \& White 1996} {Navarro-Frenk-White profile}
 {Tormen 1996}
\phrase{Eggen, Lynden-Bell \& Sandage 1962} {Monolithic collapse} {Gilmore 1996}
\phrase{Farouki \& Shapiro 1981} {Harassment} {Moore et al.\ 1996}
\phrase{White \& Rees 1978} {Halo substructure} {Moore, Katz \& Lake 1996}
\phrase{Klypin \& Shandarin 1983} {The Cosmic Web}
 {Bond, Kofman \& Pogosyan 1996}
\phrase{Lacey \& Cole 1993} {Merger trees} {Rodrigues \& Thomas 1996}
\phrase{Couchman \& Rees 1986} {End of the Dark Ages} {Gnedin \& Ostriker 1997}
\phrase{White \& Rees 1978} {Overcooling problem} {Steinmetz 1997}
\phrase{Lilly et al.\ 1996, Madau et al.\ 1996} {The Madau Plot}
 {Trentham, Blain, Goldader 1999}
\phrase{Larson, Tinsley \& Caldwell 1980} {Strangulation}
 {Balogh \& Morris 2000}
\phrase{Ferrarese \& Merritt 2000} {$M$--$\sigma$ relation}
 {Ferrarese \& Merritt 2000}
\phrase{Seljak 2000, Peacock \& Smith 2000} {The Halo Model} {White 2001}
\phrase{Flores \& Primack 1994, Moore 1994} {Cores vs.\ cusps}
 {van den Bosch 2001}
\phrase{Kormendy 1993} {Pseudo-bulges} {Kormendy 2001}
\phrase{Jing, Mo \& Boerner 1998} {Halo Occupation Distribution} {Berlind 2001}
\phrase{Cowie et al.\ 1996} {Downsizing} {Tran 2002}
\phrase{Spite \& Spite 1978} {Stellar archaeology} {Cohen et al.\ 2002}
\phrase{Katz \& White 1993, Klypin et al.\ 1999} {Substructure crisis}
 {Somerville 2002}
\phrase{Katz et al.\ 2003} {Cold streams} {Dekel \& Birnboim 2004}
\phrase{Springel, Di Matteo \& Hernquist 2005} {AGN quenching}
 {Scannapieco, Silk \& Bouwens 2005}
\phrase{van Dokkum 2005} {Dry mergers vs.\ wet mergers}
 {Naab, Jesseit, Burkert 2006}
\phrase{Bond 1993} {Gastrophysics} {Primack 2007}
\phrase{Croton et al.\ 2006} {Radio-mode feedback} {Croton et al.\ 2006}
\phrase{Gao, Springel \& White 2005} {Assembly bias}
 {Croton, Gao \& White 2007}
\phrasefoot{????}

\vfil\eject

\bigskip
\centerline{\bigbf Lessons}
\medskip
\noindent
What can we learn from this timeline by viewing it as a process?
First, if we simply count ideas, and assume there is a bandwagon effect
associated with each one, then the duration of each fad is approximately
one year [7].  The corollary to this finding, and the advice for new
researchers, is that one should
jump quickly onto bandwagons before they pass.  The second point is that not
all of the concepts listed are entirely new, and it may even seem to the
cynical reader that some apparently new suggestions are simply recycled from
earlier ones,
but with new names.  This leads to another recommendation for those who wish
to make an impact on galaxy formation -- study what is already known, find
something that has not been highlighted much before,
and come up with a new name for it.

\bigskip
\centerline{\bigbf Sidelines}
\medskip
\noindent
We end with a list of ideas that are at least a little
off the mainstream of galaxy formation research.  These are the ``also-ran''
or ``dead-end'' concepts, some of which seemed a bit outr{\'e} in the
first place, while others appeared promising briefly, but were
ultimately seen to be merely a distraction [8].
Such ideas might include:
primordial turbulence;
continuous creation;
cosmic explosions;
mock gravity;
isocurvature baryons;
cosmic string wakes;
textures;
late time phase transitions;
warm dark matter;
self-interacting dark matter;
cooling flows;
hyper-extended perturbation theory;
retarded galaxies in voids;
jet-triggered star formation;
fractal structure;
plasma cosmology;
MOND;
MOG; 
primordial black holes;
primordial magnetic fields;
etc.

No doubt there will be many more such notions to come.  Only time will
tell whether any of them become part of the main narrative.

\bigskip
\centerline{\bigbf The future}
\medskip
\noindent
Experts disagree on whether we are about to enter a golden ``precision
era'' of galaxy formation or whether the subject is essentially over,
with just the weather-prediction details left to fuss about.
Although precise future directions are unknown, some general predictions
are possible: (1) galaxy formation will not be completely ``solved''
in the near future; (2) ambitious multi-wavelength surveys will extend our
empirical understanding of the high-$z$ Universe;
(3) there will continue to be phrases spun to
describe new ideas; (4) some of these ideas will be old ideas, dressed up;
(5) some ideas will be crazy, and will fall by the wayside; but (6) some ideas
will genuinely progress the field, inspiring a new generation of galaxy
fabricators.

\bigskip
\bigskip
\centerline{\bigbf Notes and references}
\medskip
\eight
\parskip = 0pt plus 1pt minus 1pt

\nn [1]\ The boundaries are of course quite blurred between galaxy formation
and the nearby topics of star formation, cluster physics
and large-scale structure.  The choice of how far to explore around these
boundaries is necessarily quite subjective.

\nn [2]\ Using analogies from the world's of entertainment, the media and
politics.  The web 2.0 version would be ``meme''.

\nn [3]\ For tracking down the source of a phrase or idea, I have tried to
simplify a complicated history by picking a single paper in most cases.
There will undoubtedly be errors in this process, and I apologise for
either getting the originator wrong, or missing an earlier example of the
use of the phrase.  The main aim of the ``usage'' column is to trace when
the idea started to
become popular in the literature, and to be definitive I focus on abstracts
of papers in journals or conference proceedings, ignoring AAS abstracts or
telescope proposals.

\nn [4]\ One could study the longevity of each idea, and whether
specific periods of time, scientists or journals have been
more productive, etc.  This is complicated by the fact that some phrases
were originally coined with a slightly different meaning, e.g.\
``starburst'' (the nucleus only), ``quenching'' (of radio emission) and
``cold streams'' (tidal debris).  Tracking the citations is made more
difficult as a result of the natural tendency of researchers to consider
history to have started when they entered the field.  Because of these
complications
I leave it to more serious historians of science to trace the
detailed evolution of each idea.  My colleague Dr.\ Frolop has already
embarked on such a project.

\nn [5]\ Other areas of astrophysics could surely be traced in a similar
way; however, galaxy formation seems to have more than its fair share of these
catchphrases, presumably because it is a complex subject, which has to be
tackled from many different perspectives.

\nn [6]\ This term is often used in popular-level presentations, but rarely
in technical papers.

\nn [7]\ Many ideas are current at the same time of course, so this estimate
is really the new bandwagon rate.  Catchphrases differ in their longevity,
some taking a long time
to be picked up in the literature after first being discussed, while others
resonating instantly with other researchers.

\nn [8]\ Although of course no idea ever dies entirely.

\nn [9]\ This article made use of NASA's Astrophysical Data System
Bibliographic Services.
I wish to acknowledge discussions with and advice from many
colleagues, particularly those who refrained from simply pointing out
which ideas had been theirs.

\smallskip

\pp Alexander S., 1852, AJ, 2, 36

\pp Audouze J., Tinsley B.M., 1976, ARAA, 14, 43

\pp Balogh M.L., Morris S.L., 2000, MNRAS, 318, 703

\pp Bardeen J.M., 1968, AJ, 73, 164

\pp Bardeen J.M., Bond J.R., Kaiser N., Szalay A.S., 1986, ApJ, 304, 15

\pp Baugh C.M., 2006, Rept.\ Prog.\ Phys., 69, 3101

\pp Benson A.J., 2010, Phys. Rep., 495, 33

\pp Berlind A.A., 2001, Ph.D.\ thesis, Ohio State University

\pp Blumenthal G.R., Faber S.M., Primack J.R., Rees M.J., 1984, Nature, 311, 517

\pp Bond J.R., Centrella J., Szalay A.S., Wilson J.R., 1984, in ``Formation
 and evolution of galaxies and large structures in the universe, 3rd Moriond
 Astrophysics Meeting'', Reidel, Dordrecht, p.$\,87$

\pp Bond J.R., Kofman L., Pogosyan D., 1996, Nature, 380, 603

\pp Bond J.R., 1993, in ``The environment and evolution of galaxies, 3rd Tetons
 summer school'', ed.\ J.M. Shull, H.A. Thronson, Kluwer, Dordrecht, p.$\,3$

\pp Bonnor W.B., 1957, MNRAS, 117, 104

\pp Brosche P., 1970, A\&A, 6, 240

\pp Burbidge G.R., Burbidge E.M., 1959, ApJ, 130, 15

\pp Burbidge E.M., Burbidge G.R.., Fowler W.A., Hoyle F., 1957, Rev.\ Mod.
 Phys., 29, 547

\pp Chevalier R.A., Clegg A.W., 1986, Nature, 317, 44

\pp Cohen J.G., et al., 2002, AJ, 124, 470

\pp Couchman H.M.P., Rees M.J., 1986, MNRAS, 221, 53

\pp Cowie L.L., Songaila A., Hu E., Cohen J.G., 1996, AJ, 112, 839

\pp Croton D.J.,  et al., 2006, MNRAS, 365, 11

\pp Croton D.J., Gao L., White S.D.M., 2007, MNRAS, 374, 1303

\pp Crommelin A.C.D., 1918, JRASC, 12, 33

\pp Curtis H.D., 1917, PASP, 29, 206

\pp Dekel A., Birnboim Y., 2004, in ``The new cosmology: conference on
 strings and cosmology'', AIP Conf.\ Proc., Vol.\ 743, p.$\,162$

\pp Dekel A., Kowitt M., Shaham J., 1981, ApJ, 250, 561

\pp Dekel A., Silk J., 1986, ApJ, 303, 39

\pp de Vaucouleurs G., Olson D.W., 1982, ApJ, 256, 346

\pp Djorgovski S., 1987, in ``Structure and dynamics of elliptical galaxies'',
 Reidel, Dordrecht, p.$\,79$

\pp Doroshkevich A.G., 1970, Astrophysica, 6, 320

\pp Doroshkevich A.G., Zeldovich Ya.B., 1975, Ap\&SS, 35, 55

\pp Doroshkevich A.G., Zeldovich Ya.B., Novikov I.D., 1967, Sov. Astron.,
 11, 233

\pp Dressler A., 1980, ApJ, 236, 351

\pp Dressler A., Lynden-Bell D., Burstein D., Davies R.L., Faber S.M.,
 Terlevich R.J., Wegner G., 1987, ApJ, 313, 42

\pp Dunlop J.S., Guiderdoni B., Rocca-Volmerange B., Peacock J.A.,
 Longair M.S., 1989, MNRAS, 240, 257

\pp Earl of Rosse, 1850, Phil.\ Trans.\ Royal Soc., 140, 499

\pp Efstathiou G., Fall S.M., Hogan C., 1979, MNRAS, 189, 203

\pp Efstathiou G., Silk J., 1983, Fund. Cosm. Phys., 9, 1

\pp Eggen O.J., Lyden-Bell D., Sandage A.R., 1962, ApJ, 136, 748

\pp Einasto J., Kaasik A., Saar E., 1974, Nature, 250, 309

\pp Evrard A.E., 1989, ApJ, 341, 26

\pp Faber S.M., Jackson R.E., 1976, ApJ, 204, 668

\pp Farouki R., Shapiro S.L., 1981, ApJ, 243, 32

\pp Ferrarese L., Merritt D., 2000, ApJ, 539, L9

\pp Flores R.A., Primack J.R., 1994, ApJ, 427, L1

\pp Franco J., Cox D.P., 1983, ApJ, 273, 243

\pp Gamow 1948, Phys.\ Rev., 74, 505

\pp Gamow 1953, AJ, 58, 39

\pp Gao L., Springel V., White S.D.M., 2005, MNRAS, 363, L66

\pp Gilmore G., 1996, ASP Conf.\ Ser., Vol.\ 92, p.$\,161$

\pp Gingold R.A., Monaghan J.J., 1977, MNRAS, 181, 375

\pp Gingold R.A., Monaghan J.J., 1981, MNRAS, 197, 461

\pp Giovanelli R., Haynes M.P., Chincarini G.L., 1981, ApJ, 247, 383

\pp Gnedin N.Y., Ostriker J.P., 1997, ApJ, 486, 581

\pp Gold T., Hoyle F., 1959, in ``Paris Symposium on Radio Astronomy'',
ed. R.M. Bracewell, Stanford University Press, Palo Alto, p.$\,$583

\pp Gott J.R., 1973, ApJ, 186, 481

\pp Gott J.R., 1975, ApJ, 201, 296

\pp Gott J.R., 1977, ARA\&A, 15, 235


\pp Gott J.R., Turner E.L., 1977, ApJ, 213, 309

\pp Gunn J.E., Gott J.R. 1972, ApJ, 176, 1

\pp Gunn J.E., Peterson B.A., 1965, ApJ, 142, 1633

\pp Harrison E.R., 1970, MNRAS, 148, 119

\pp Harrison E.R., 1971, MNRAS, 154, 167

\pp Heckman T.M., Armus L., Miley G.K., 1987, AJ, 93, 276

\pp H{\'e}non M., 1964, Ann.\ d'Astrophys., 27, 83

\pp Hernquist L., Spergel D.N., 1992, ApJ, 399, L117

\pp Herschel W., 1786, Phil. Trans. R. Soc. London, 76, 457

\pp Holmberg E., 1941, ApJ, 94, 385

\pp Hoyle F., 1951, in ``Problems of Cosmical Aerodynamics'', Central Air
 Documents Office, Dayton, Ohio, p.$\,$195

\pp Hoyle F., 1953, ApJ, 118, 513

\pp Hubble E.P., 1923, Popular Astron., 31, 644

\pp Hubble E.P., 1926, ApJ, 64, 321

\pp Hubble E.P., 1936, ``The Realm of the Nebulae'', Yale University Press,
 New Haven

\pp Humason M.L., 1936, ApJ, 83, 10

\pp Jeans J.H., 1902, Phil. Trans. R. Soc. London A, 199, 1

\pp Jing Y.P., Mo H.J., Boerner G., 1998, ApJ, 494, 1

\pp Jones B.J.T., 1976, Rev. Mod. Phys., 48, 107

\pp Jones B.J.T., Palmer P.L., 1985, in `Galaxies, axisymmetric systems and
 relativity', Cambridge University Press, Cambridge, p.$\,3$

\pp Kaiser N., 1984, ApJ, 284, L9

\pp Kant I., 1755, Allgemeine Naturgeschichte und Theorie des Himmels,
Johann Friedrich Petersen, K{\"o}nigsberg and Leipzig

\pp Katz N., 1992, PASP, 104, 852

\pp Katz N., Kere{\v s} D., Dav{\'e} R., Weinberg D.H., 2003, in ``The
 IGM/galaxy connection'', ASSL Conf.\ Proc., Vol.\ 281, Kluwer, Dordrecht,
 p.$\,185$

\pp Katz N., White S.D.M., 1993, ApJ, 412, 455

\pp Kauffmann G., White S.D.M., Guiderdoni B., 1993, MNRAS, 264, 201

\pp Klypin A.A., Kravtsov A.V., Valenzuela O., Prada F., 1999, ApJ, 522, 82

\pp Klypin A.A., Shandarin S.F., 1983, MNRAS, 204, 891

\pp Kormendy J., 1979, ApJ, 227, 714

\pp Kormendy J., 1993, in ``Galactic Bulges'', IAU Symp.\ 153, ed.\
 H. Dejonghe, H.J. Habing, Kluwer, Dordrecht, p.$\,209$

\pp Kormendy J., 2001, in ``Galaxy Disks and Disk Galaxies'', ASP Conf.\ Ser.,
 Vol.\ 230, p.$\,247$

\pp Lacey C., Cole S., 1993, MNRAS, 262, 627

\pp Larson R.B., 1969, MNRAS, 145, 405

\pp Larson R.B., Tinsley B.M., Caldwell C.N., 1980, ApJ, 237, 692

\pp Layzer D., 1963, ApJ, 137, 351

\pp Layzer D., 1964, ARA\&A, 2, 341

\pp Leir A.A., van den Bergh S., 1977, ApJS, 34, 381

\pp Lifshitz E.M., 1946, J. Phys. U.S.S.R., 10, 110

\pp Lilly S.J., Le Fevre O., Hammer F., Crampton D., 1996, ApJ, 460, L1

\pp Longair M.S., 1998, 2008, ``Galaxy formation'', Springer-Verlag, Berlin

\pp Lynden-Bell D., 1967, MNRAS, 136, 101

\pp Madau P., Ferguson H.C, Dickinson M.E., Giavalisco M., Steidel C.C.,
 Fruchter A., 1996, MNRAS, 283, 1388

\pp Mathews W.G., Baker J.C., 1971, ApJ, 170, 241

\pp Messier C., 1781, Connoissance des Temps, pp.$\,$227--267 (published
 in 1784)

\pp Mo H., Van den Bosch F., White S.D.M., 2010, ``Galaxy Formation and
 Evolution'', Cambridge University Press, Cambridge

\pp Moore B., 1994, Nature, 370, 629

\pp Moore B., Katz N., Lake G., 1996, ApJ, 457, 455

\pp Moore B., Katz N., Lake G., Dressler A., Oemler A., 1996, Nature, 379, 613

\pp Naab T., Jesseit R., Burkert A., 2006, MNRAS, 372, 839

\pp Navarro J.F., Frank C.S., White S.D.M., 1996, ApJ, 462, 563

\pp Norman C., Silk J., 1979, ApJ, 233, L1

\pp Novikov I.D., 1975, AZh, 52, 1038

\pp Ostriker J.P., Hausman M.A., 1977, ApJ, 217, L125

\pp Ostriker J.P., Peebles, 1973, ApJ, 186, 467

\pp Ostriker J.P., Tremaine S., 1975, ApJ, 202, L113

\pp Ozernoy L.M., 1973, Astron. Tsirk, 804, 1

\pp Padmanabhan T., Vasanthi M.M., 1985, JApA, 6, 261

\pp Park C., 1991, MNRAS, 251, 167

\pp Peacock J.A., Smith R.E., 2000, MNRAS, 318, 1144

\pp Peebles P.J.E., 1965, ApJ, 142, 1317

\pp Peebles P.J.E., 1969, ApJ, 155, 393

\pp Peebles P.J.E., 1974, I.A.U. Symp., 58, 55

\pp Peebles P.J.E., 1984, in `Clusters and groups of galaxies', ed.\
 F. Mardirossian et al., Reidel, Dordrecht, p.$\,405$

\pp Peebles P.J.E., Dicke R.H., 1968, ApJ, 154, 891

\pp Perrenod S.C., 1978, ApJ, 224, 285

\pp Polyachenko V.L., Fridman A.M., 1976, ``Equilibrium and stability of
gravitating systems'', Nauka, Mosow

\pp Postman M., Geller M.J., 1984, ApJ, 282, 95

\pp Press W.H., Schechter P., 1974, ApJ, 187, 425

\pp Primack J.R., 2007, Nucl.\ Phys. B Supp., 173, 1

\pp Primack J.R., Blumenthal G.R., 1984, in ``Formation and evolution of
 galaxies and large structures in the universe, 3rd Moriond Astrophysics
 Meeting'', Reidel, Dordrecht, p.$\,163$

\pp Quinn P.J., in ``Nearly normal galaxies: From the Planck time to the
 present'', ed.\ S.M. Faber, Springer-Verlag, New York, p.$\,138$

\pp Rawson-Harris D., 1969, MNRAS, 143, 49

\pp Rees M.J., 1978, in ``Observational cosmology, 8th Saas-Fee Advanced
 Course'', Sauveny, Geneva Observatory, p.$\,259$

\pp Rieke G.H., 1982, in ``Extragalactic radio sources'', Reidel, Dordrecht,
 p.$\,239$

\pp Rieke G.H., Lebofsky M.J., 1979, ARAA, 17, 477

\pp Roberts I., 1895, MNRAS, 56, 70

\pp Rodrigues D.D.C., Thomas P.A., 1996, in ``Formation of the Galactic Halo
 \dots Inside and Out'', ASP Conf.\ Ser., Vol.\ 92, p.$\,505$

\pp Salpeter E.E., 1964, ApJ, 140, 796

\pp Sandage A., Tammann G.A., 1976, ApJ, 210, 7

\pp Scannapieco E., Silk J., Bouwens R., 2005, ApJ, 635, L13

\pp Schaeffer R., Silk J., 1984, A\&A, 130, 131

\pp Schwarzschild M., Spitzer L., 1953, Obs., 73, 77

\pp Sciama D.W., 1955, MNRAS, 115, 3

\pp Searle L., Zinn R., 1978, ApJ, 225, 357

\pp Seljak U., 2000, MNRAS, 318, 203

\pp Sofue Y., 1969, PASJ, 21, 211

\pp Somerville R.S., 2002, ApJ, 572, L23

\pp Spinrad H., 1993, in ``The environment and evolution of galaxies, 3rd
 Tetons summer school'', ed.\ J.M. Shull, H.A. Thronson, Kluwer, Dordrecht,
 p.$\,151$

\pp Spinrad H., 2005, ``Galaxy Formation and Evolution'', Springer-Verlag,
 Berlin

\pp Spite F., Spite M., 1978, A\&A, 67, 23

\pp Spitzer L., 1942, ApJ, 95, 329

\pp Springel V., Di Matteo T., Hernquist L., 2005, ApJ, 620, L79

\pp Steinmetz M., 1997, in ``The Early Universe with the VLT'', ed.\
 J. Bergeron, Springer, Berlin, p.$\,156$

\pp Str{\"o}mberg G., 1934a, ApJ, 79, 460

\pp Str{\"o}mberg G., 1934b, ApJ, 80, 327

\pp Sunyaev R.A., Zeldovich Ya.B., 1972, A\&A, 20, 189

\pp Thuan T.X., 1975, Nature, 257, 774

\pp Toomre A., 1974, I.A.U. Symp., 58, 347

\pp Toomre A., Toomre J., 1972, ApJ, 178, 623

\pp Tormen G., 1996, in ``Mapping, measuring, and modelling the universe'',
 ASP Conf.\ Ser., Vol.\ 94, p.$\,131$

\pp Tran K.-V.H., 2002, Ph.D.\ thesis, U.C. Santa Cruz

\pp Trentham N., Blain A.W., Goldader J., 1999, MNRAS, 305, 61

\pp Tully R.B., Fisher J.R., 1976, A\&A, 54, 661

\pp van Albada G.B., 1961, AJ, 66, 590

\pp van den Bergh S., 1979, Astron. Nachr., 300, 225

\pp van den Bosch F.C., Swaters R.A., 2001, MNRAS, 325, 1017

\pp van Dokkum P.G., 2005, AJ, 130, 2647

\pp von Hoerner S., 1960, Zeit.\ f{\"u}r Astrophysik, 50, 184

\pp von Humboldt, 1845, ``Kosmos'', Vol.\ I, Hippolyte Bailli{\`e}re, London

\pp Wagner R.L., 1978, A\&A, 62, 9

\pp White M., 2001, MNRAS, 321, 1

\pp White S.D.M., 1995, in ``1993 Les Houches Lectures on Large-Scale
 Structure'', ed.~R. Schaeffer, Elsevier, Netherlands

\pp White S.D.M., Frenk C., 1991, ApJ, 457, 645

\pp White S.D.M., Rees M.J., 1978, MNRAS, 183, 341

\pp Zeldovich Ya.B., 1970a, Astrofizika, 6, 319

\pp Zeldovich Ya.B., 1970b, A\&A, 5, 84

\pp Zwicky F., 1933, Helv.\ Phys.\ Acta, 6, 110

\pp Zwicky F., 1937, ApJ, 86, 217

\bye